\documentclass[12pt,letter]{article}
\pdfoutput=1
\usepackage{graphicx, epsfig, color,cite}
\usepackage{amsmath}
\usepackage{amssymb}
\usepackage{caption,subcaption,graphicx}
\usepackage[hidelinks]{hyperref}
\def\eslt{\not\!\!\!{E_T}}
\def\to{\rightarrow}

\def\bi{\begin{itemize}}
\def\ei{\end{itemize}}
\def\te{\tilde e}

\def\tchi{\tilde\chi}

\def\tu{\tilde u}

\def\tb{\tilde b}

\def\tst{\tilde t}
\def\ttau{\tilde \tau}

\def\tg{\tilde g}
\def\tnu{\tilde\nu}

\def\tw{\widetilde\chi^{\pm}}
\def\twns{\widetilde\chi}
\def\tz{\widetilde\chi^0}
\def\alt{\lesssim}
\def\agt{\gtrsim}
\def\be{\begin{equation}}  
\def\ee{\end{equation}}  
\def\bea{\begin{eqnarray}}  
\def\eea{\end{eqnarray}}  

\newcommand\snowmass{\begin{center}\rule[-0.2in]{\hsize}{0.01in}\\\rule{\hsize}{0.01in}\\
\vskip 0.1in Submitted to the  Proceedings of the US Community Study\\ 
on the Future of Particle Physics (Snowmass 2021)\\ 
\rule{\hsize}{0.01in}\\\rule[+0.2in]{\hsize}{0.01in} \end{center}}

\begin{document}
\begin{titlepage}
\begin{flushright}
  {UH-511-1324-22}
\end{flushright}

\vspace{0.5cm}
\begin{center}
\renewcommand{\thefootnote}{\fnsymbol{footnote}}
{\Large \bf Angular cuts to reduce the $\tau\bar{\tau} j$ background to the higgsino signal at the LHC}\\
\vspace{1.2cm}
{\large Howard Baer$^1$\footnote[2]{Email: baer@ou.edu },
Vernon Barger$^2$\footnote[3]{Email: barger@pheno.wisc.edu},
Dibyashree Sengupta$^3$\footnote[4]{Email: dsengupta@phys.ntu.edu.tw} 
and Xerxes Tata$^4$\footnote[5]{Email: tata@phys.hawaii.edu}
}\\ 
\vspace{1.2cm} \renewcommand{\thefootnote}{\arabic{footnote}}
{\it 
$^1$Homer L. Dodge Department of Physics and Astronomy,\\
University of Oklahoma, Norman, OK 73019, USA \\[3pt]
}
{\it 
$^2$Department of Physics,
University of Wisconsin, Madison, WI 53706 USA \\[3pt]
}
{\it 
$^3$Department of Physics,
National Taiwan University, Taipei, Taiwan 10617, R.O.C. \\[3pt]
}
{\it 
$^4$Department of Physics and Astronomy,
University of Hawaii, Honolulu, HI, USA\\[3pt]
}

\end{center}

\vspace{0.5cm}
\begin{abstract}
\noindent
We re-examine higgsino pair production in association with a hard QCD
jet at the LHC. We focus on $\ell^+\ell^-+\eslt+j$ events from the
production and subsequent decay, $\tchi_2^0\to\tchi_1^0\ell^+\ell^-$, of
the heavier neutral higgsino.  The novel feature of our analysis is that
we propose angular cuts to reduce the irreducible background from
$Z(\to\tau\bar{\tau})+j$ events more efficiently than the $m_{\tau\tau}^2<0$
cut that has been used by the ATLAS and CMS collaborations. Additional
cuts, needed to reduce backgrounds from $t\bar{t}$, $WWj$ and
$W/Z+\ell\bar{\ell}$ production, are also delineated.  We evaluate the
reach of LHC14 for 300 and 3000~fb$^{-1}$ and stress that the dilepton
mass distribution would serve to characterize the higgsino signal.

\end{abstract}
\snowmass
\end{titlepage}
\begin{center}
\underline{{\bf Executive Summary}}
\end{center}

The ATLAS and CMS collaborations have been searching for light higgsinos
with a compressed spectrum -- whose existence is, perhaps, the most
robust prediction of natural SUSY -- via monojet events with an
additional soft $e^+e^-$ or $\mu^+\mu^-$ pair, coming mainly from the
decay $\chi_2^0\to \chi_1^0 \ell\bar{\ell}$.  $Z(\to\tau\bar{\tau})+ j$
production (where both taus decay leptonically) is an important
irreducible SM background to the higgsino signal. This background can be
considerably reduced by requiring that $m_{\tau\tau}^2<0$, where
$m_{\tau\tau}^2$ is constructed in the approximation (valid for
relativistic taus from $Z$ decays) that the tau decay products are
collinear with the parent tau direction.  {\bf We have devised angular
  cuts that reduce the tau pair plus jet background much more
  efficiently than this di-tau mass cut.}

Our results are exhibited in the Table below for three SUSY benchmark
points, BM1, BM2 and BM3 with a higgsino masses around 150, 200 and
300~GeV and $\Delta m \sim 12, 16$ and 4.3~GeV, respectively. Also shown
are the important SM backgrounds.  We see that while the di-tau mass cut
reduces the tau background by about a factor of 4 (row 2), the angle
cuts reduce this by a factor of 50 (row 3) with a relatively small loss
in signal efficiency. The last line shows the signal and the background
after additional cuts necessary to reduce the other backgrounds.
\begin{table}[h!]
\centering
\begin{tabular}{lcccccccc}
\hline
cuts/process & $BM1$ & $BM2$ & $BM3$ & $\tau\bar{\tau}j$ & 
$t\bar{t}$ & $WWj$ & $W\ell\bar{\ell}j$ & $Z\ell\bar{\ell}j$ \\
\hline
Basic cuts       & 1.2 & 0.19 & 0.07 & 94.2 & 179 & 35.9 & 14.7 & 5.9 \\
Basic+$m_{\tau\tau}^2<0$        & 0.92 & 0.13 & 0.043 & 23.1 & 75.6 & 12.8 & 7.7 
& 3.2 \\
Basic+angle        & 0.68 & 0.12 & 0.04 & 1.8 & 130 & 22 & 11.0 & 4.9 \\
Additional cuts        & 0.25 & 0.032 & 0.017 & 0.088 & 0.29 & 0.39 & 0.15 & 0.07 \\
\hline
\end{tabular}
\caption*{Table: Cross sections (in $fb$) for signal benchmark points and the
 various SM backgrounds. }
\end{table}

Our projected reach for the signal at the LHC with 300~fb$^{-1}$ and for
3000~fb$^{-1}$ is shown by dashed lines in the figure, and compared with
corresponding projections for the HL-LHC obtained by the ATLAS and CMS
collaborations.

\begin{figure}[!bhtp]
\begin{center}
\includegraphics[height=0.2\textheight]{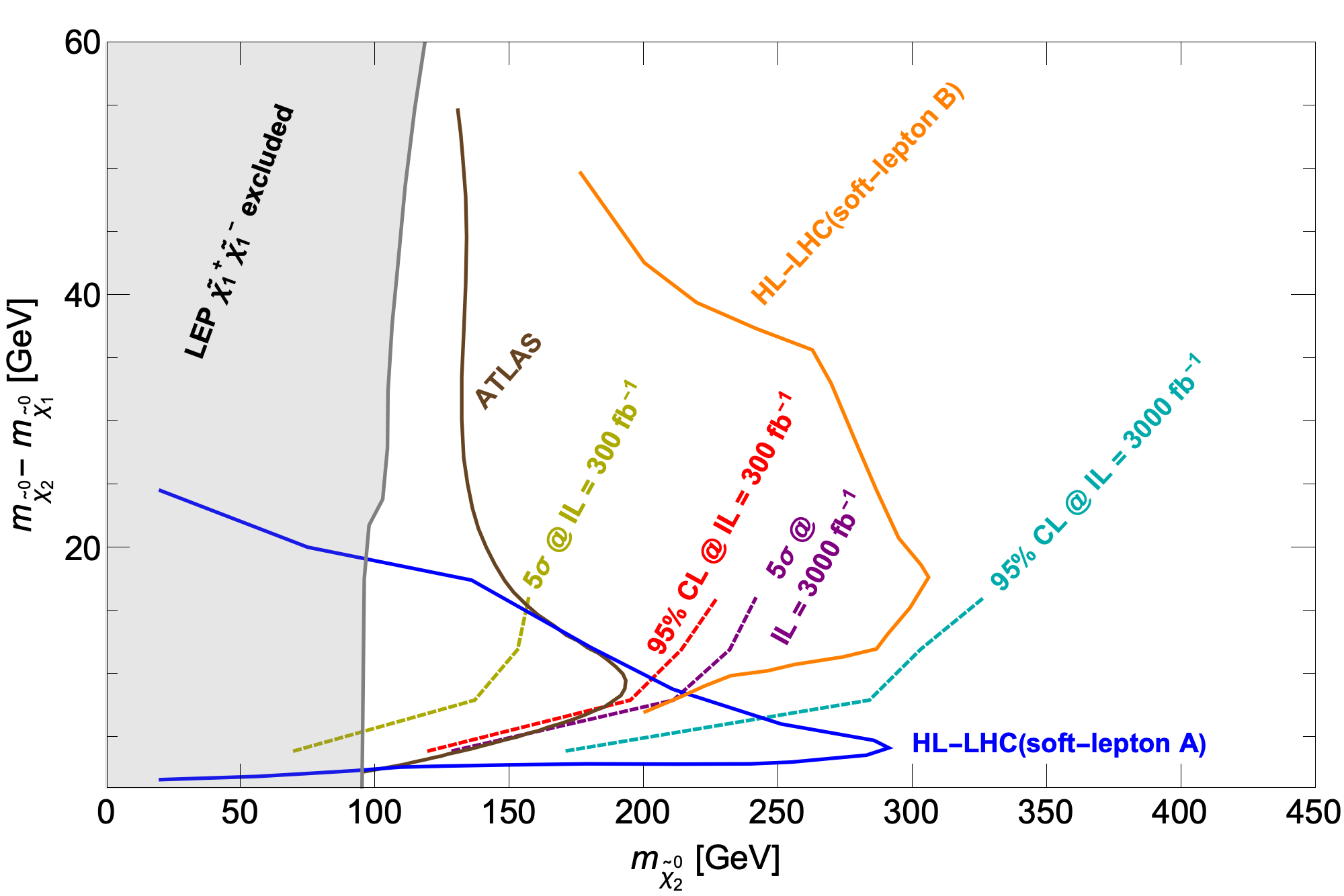}
\caption*{Figure: Our projections for the LHC reach (dashed lines) along with the
  current 95\% CL exclusion (ATLAS) and the projected 95\% CL exclusions
  from two different analyses for the HL-LHC.}
\end{center}
\end{figure}

\newpage

\section{Introduction}
\label{sec:intro}

It is generally expected that higgsinos cannot be much heavier than the
electroweak scale if weak scale supersymmetry is the new physics that
stabilizes the Higgs sector \cite{smallmu}.  The precise upper limit on
the higgsino mass depends on the degree of fine-tuning that is
considered acceptable: for no worse than a part in thirty electro-weak
fine-tuning ($\Delta_{\rm EW} < 30$) \cite{ewft}, higgsinos are expected
to be lighter than $\sim 350$~GeV, while all other super-partners could
be in the multi-TeV range \cite{ewft,heavy}.  For this reason, the
search for light higgsinos has become an important part of the
experimental SUSY program at the LHC \cite{atlas,cms}.

In the generic situation of natural SUSY models where electroweak
gauginos are much heavier than the higgsinos (gauginos with multi-TeV
masses do not destabilize the Higgs vacuum-expectation-value)
the mass splitting between the two lightest neutral higgsinos
is very small-- $\sim$4-25 GeV-- so that the visible products of
higgsino decays to the (higgsino-like) LSP\footnote{ We assume
  conservation of $R$-parity and also note that the lightest, thermally
  produced higgsino comprises only a portion of the observed relic cold
  dark matter abundance.}  are correspondingly soft.  As a result,
although higgsino pair production cross sections at LHC14 are
substantial ($\sim 100-1000~fb$ for higgsino masses in the 120-350~GeV
range), the visible decay products from higgsino pair production may be
hidden under enormous backgrounds from Standard Model (SM)
processes\cite{bbh}.  Because the visible decay products of the
higgsinos are-- for practical purposes-- invisible, this led to the
suggestion that it might be possible to search for higgsino pairs
produced in association with a hard object $X$, where $X$ is a
mono-jet\cite{monojet}, mono-photon\cite{monophoton}, mono-gauge
boson\cite{monoZ} or mono-Higgs boson\cite{monoH}.  While it is possible
to gain a statistical significance of the signal exceeding $5\sigma$ by
requiring a hard enough cut on the mono-jet, the signal to background
ratio is at the $\sim 1\%$ level, so that this strategy is viable only
if the backgrounds can be controlled to percent level or below
precision.

An alternative strategy is to detect the {\em soft leptons} produced via
higgsino decays in higgsino pair events triggered by a hard monojet (or
the $\ \eslt$) to further reduce the SM background
\cite{soft,kribs,bmt,hkmp}. Dileptons from $\tz_2$ decays would
necessarily be of opposite sign (OS) and the same flavour (SF), and
further, their invariant mass is kinematically bounded above by $\Delta
m \equiv m_{\tz_2}-m_{\tz_1}$. The higgsino mono-jet plus soft dilepton
signal has been studied by several groups, and has been used by both the
ATLAS \cite{atlas} and CMS \cite{cms} collaborations for their searches
for higgsinos with a compressed spectrum. The main purpose of this note
is to point out an analysis strategy involving {\em angular cuts} that is
more effective than the currently used method that uses the di-tau
invariant mass to beat down the important SM background to the
higgsino signal arising from tau pair production in association with a
hard jet (the leptons come from the decays of the taus). We also
delineate other cuts that reduce the other backgrounds, and map out
the reach of the HL-LHC in the $m_{\tz_2}$ vs. $\Delta m$ mass
plane. This Snowmass 2021 contribution summarizes and streamlines the results of
Ref.~\cite{angular} with an emphasis on the angular cuts to reduce the
important irreducible background from $Z\to \tau\bar{\tau}$ events.

\section{Higgsino Signal and SM background processes} \label{sec:processes}

We focus on the signal from higgsino pair production in association with
a QCD jet at the LHC: $pp \to \tchi_1^+\tchi_1^-,\ \tz_1\tz_2,\ \tw_1\tz_2 +j$,
where $\tw_1\to \ell\nu\tz_1$ and $\tz_2\to \ell\bar{\ell}\tz_1$
($\ell= e, \mu$). Requiring a hard jet boosts the lepton daughters of the
higgsino in addition to providing a trigger for the events. We evaluate
SM backgrounds to the monojet plus OS/SF dilepton higgsino signal from:
\begin{itemize}
\item $\tau\bar{\tau}j$ production,
\item $t\bar{t}$ production,
\item $WWj$ production,
\item $W\ell\bar{\ell}j$ production, and
\item $Z\ell\bar{\ell}j$ production
\ei 
in the SM.

For our calculations, we use {\sc MadGraph}~2.5.5~\cite{madgraph}
interfaced to {\sc PYTHIA} v8~\cite{pythia} via the default
MadGraph/PYTHIA interface with default parameters for showering and
hadronization to generate $pp$ collision events with $\sqrt{s}=14$~TeV.
Detector simulation is performed by {\sc Delphes} using the default
Delphes~3.4.2~\cite{delphes} ``ATLAS'' parameter card.
We adopt the anti-$k_T$ jet algorithm~\cite{Cacciari:2008gp} with $R =
0.6$, the default value in the ATLAS Delphes card, to form jets.  Jet
finding in Delphes is implemented via {\sc
  FastJet}~\cite{Cacciari:2011ma}.  We consider only jets with
transverse energy satisfying $E_T(jet) > 40$ GeV and pseudorapidity
satisfying $|\eta(jet)| < 3.0$ in our analysis.  We implement the
default Delphes $b$-jet tagger and implement a $b$-tag efficiency of
85\%~\cite{ATLAS:2015dex}. We identify leptons with $E_T> 5$~GeV and
within $|\eta (\ell )| < 2.5$. We label them as {\em isolated leptons}
if the sum of the transverse energy of all other objects (tracks,
calorimeter towers, etc.) within $\Delta R = 0.5$ of the lepton
candidate is less than $10\%$ of the lepton $E_T$.

In order to assess the level of the SUSY signal relative to the SM
backgrounds, we have selected three benchmark points with varying
higgsino masses and varying mass gaps between the heavier higgsinos and
the LSP: the higgsino production rate obviously depends on the higgsino
mass, and the lepton detection efficiency is very sensitive to the
higgsino mass gap, $\Delta m$. We use the
computer code Isajet 7.88\cite{isajet} to generate sparticle masses. The
resulting SUSY Les Houches Accord (SLHA) files are then used as inputs
to Madgraph/Pythia/Delphes for event generation.

The first two benchmark points, BM1 and BM2, are obtained from the
two-extra-parameter non-universal Higgs model (NUHM2) with parameters
$m_0,\ m_{1/2},\ A_0,\ \tan\beta,\ \mu,\ m_A$ and have higgsino masses
around 150~GeV and 300~GeV, respectively, and corresponding mass gaps
$\Delta m=12$~GeV and 16~GeV. Although not critical for a discussion of
the phenomenology of the signal, we note that both these benchmark
points satisfy our naturalness criterion, $\Delta_{\rm EW} < 30$. The
third point, labeled as BM3 (GMM$^\prime$) has higgsino masses around
200~GeV and a small $\Delta m \simeq 4.3$~GeV. Since NUHM2 models with
$\Delta_{\rm EW} < 30$ necessarily have $\Delta m \agt 10$~GeV, we have
used the natural generalized mirage mediation model\cite{ngmm}, where
$\mu$ is an input, to generate a benchmark point with this small mass
gap; this, of course, makes the signal search a challenge even though
the higgsinos are not particularly heavy.  The input parameters and SUSY
spectra from the benchmark points are listed in Table \ref{tab:bm},
along with some low energy and dark-matter-related observables along
with the degree of electroweak fine-tuning, which we view as a
conservative measure of naturalness.
\begin{table}\centering
\begin{tabular}{lccc}
\hline
parameter & $BM1$ & $BM2$ & $BM3\ (GMM^\prime)$\\
\hline
$m_0$        & 5000 & 5000 & $\textendash$ \\
$m_{1/2}$     & 1001 & 1000 & $\textendash$ \\
$A_0$        & -8000 & -8000 & $\textendash$ \\
$\tan\beta$  & 10 & 10 & 10 \\
$m_{3/2}$     & $\textendash$ & $\textendash$ & 75000  \\
$\alpha$     & $\textendash$ & $\textendash$ & 4 \\
$c_m$        & $\textendash$ & $\textendash$ & 6.9 \\
$c_{m3}$      & $\textendash$ & $\textendash$ & 6.9 \\
$a_3$        & $\textendash$ & $\textendash$ & 5.1 \\
\hline
$\mu$          & 150   & 300 & 200 \\
$m_A$          & 2000  & 2000 & 2000 \\
\hline
$m_{\tg}$   & 2425.4  & 2422.6 & 2837.3 \\
$m_{\tu_L}$ & 5295.9 & 5295.1 & 5244.6 \\
$m_{\tu_R}$ & 5427.8 & 5426.5 & 5378.0 \\
$m_{\te_R}$ & 4823.7  & 4824.5 & 4813.2 \\
$m_{\tst_1}$ & 1571.7 & 1578.4 & 1386.9  \\
$m_{\tst_2}$ & 3772.0 & 3773.0 & 3716.7 \\
$m_{\tb_1}$ & 3806.7 & 3807.6 & 3757.8 \\
$m_{\tb_2}$ & 5161.2 & 5160.2 & 5107.7 \\
$m_{\ttau_1}$ & 4746.8 & 4747.5 & 4729.8 \\
$m_{\ttau_2}$ & 5088.6 & 5088.2 & 5075.7 \\
$m_{\tnu_{\tau}}$ & 5095.4 & 5095.0 & 5084.8 \\
$m_{\tw_2}$ & -857.1 & -857.6 & -1801.9 \\
$m_{\tw_1}$ & -156.6 & -311.6 & -211.1 \\
$m_{\tz_4}$ & -869.0 & -869.8 & -1809.3 \\ 
$m_{\tz_3}$ & -451.3 & -454.7 & -1554.4 \\ 
$m_{\tz_2}$ & 157.6 & 310.1 & 207.0 \\ 
$m_{\tz_1}$ & -145.4 & -293.7 & -202.7 \\ 
$m_h$       & 124.5 & 124.6 & 125.4 \\ 
\hline
$\Omega_{\tz_1}^{std}h^2$ & 0.007 & 0.023 & 0.009  \\
$BF(b\to s\gamma)\times 10^4$ & 3.1 & 3.1 & 3.1 \\
$BF(B_s\to \mu^+\mu^-)\times 10^9$ & 3.8 & 3.8 & 3.8\\
$\Delta_{\rm EW}$ & 13.9 & 21.7 & 26.0 \\
\hline
\end{tabular}
\caption{Input parameters and sparticle masses in~GeV units for the two
  NUHM2 model benchmark points (BM1 and BM2) and one natural mirage
  mediation SUSY benchmark point (BM3 (GMM')) introduced in the text.
  We take $m_t=173.2$ GeV. The input parameters for the
  natural(generalized) mirage mediation model such as $\alpha$ and $c_m$
  have been calculated from $m_0^{MM}$ and $m_{1/2}^{MM}$ which are
  taken equal to the corresponding NUHM2 model values of $m_0$ and
  $m_{1/2}$, respectively.  The $c_m$ and $c_{m3}$ have been taken equal
  to each other so that masses of first/second and third generation
  sfermions are equal at the GUT scale so as to also match the NUHM2
  models in the second and third columns of the table.  }
\label{tab:bm}
\end{table}

\section{Higgsino Signal Analysis} \label{sec:signal}


For the SUSY signal from higgsinos, we generate events from the
reactions $pp\to\twns_1^\pm\tz_2 j$, $\tz_1\tz_2 j$ and
$\twns_1^+\twns_1^- j$, where $\tz_2\to \tz_1\ell^+\ell^-$ and $\tw_1
\to\ell\nu\tz_1$.  The dilepton plus jet signal together with the cross
sections from the various backgrounds listed in
Sec.~\ref{sec:processes}, after a series of cuts described below, is
shown in Table \ref{tab:xsec}.  The first entry labeled $BC$ (for {\it
  before cuts}) actually has parton level cuts implemented at the
Madgraph level. These cuts serve to regularaize subprocesses that are
otherwise divergent. Also, for the backgrounds with a hard QCD ISR
(labeled as $j$ in the table header), we require $p_T(j)>80$ GeV to
efficiently generate events with a hard jet. For the backgrounds
including $\gamma^*,Z^*\to\ell\bar{\ell}$ ($\ell=e$ or $\mu$), we
require $m(\ell\bar{\ell})>1$ GeV. We also require $p_T(\ell )>1$ GeV
and $\Delta R(\ell\bar{\ell})>0.01$, again at the parton level.  The $W$
daughters of top quarks in $t\bar{t}$ events are forced to decay
leptonically (into $e$, $\mu$ or $\tau$), but not so the $W$-bosons in
first entry of the $WWj$ column.  These parton events are then fed into
PYTHIA and analysed using the DELPHES detector simulation.  The leading
order cross sections (in $fb$), for both the signal as well as for the
backgrounds, are listed in row 1, labelled $BC$. At this stage, the
signal is dwarfed by SM backgrounds.

To select out signal events, we require:
\bi
\item two opposite sign, same flavour (OS/SF) isolated leptons
  with $p_T(\ell )>5$ GeV, $|\eta (\ell )|<2.5$,
\item there be at least one jet in the event, {\it i.e.}, $n_j\ge 1$
  with $p_T(j_1)>100$ GeV for identified calorimeter jets,
\item $\Delta R(\ell\bar{\ell})>0.05$ (for $\ell =e$ or $\mu$),
\item $\eslt >100$ GeV, and
\item a veto of tagged $b$-jets, $n$($b$-jet)=0.
\ei
The cross sections after this set of cuts, labeled as {\bf C1} are shown
in the next row of Table~\ref{tab:xsec}.
\begin{table}\centering
\begin{tabular}{lcccccccc}
\hline
cuts/process & $BM1$ & $BM2$ & $BM3 (GMM^\prime)$ & $\tau\bar{\tau}j$ & 
$t\bar{t}$ & $WWj$ & $W\ell\bar{\ell}j$ & $Z\ell\bar{\ell}j$ \\
\hline
$BC$        & 83.1 & 9.3 & 31.3 & 43800.0 & 41400 & 9860 & 1150.0 & 311 \\
$C1$        & 1.2 & 0.19 & 0.07 & 94.2 & 179 & 35.9 & 14.7 & 5.9 \\
$C1+m_{\tau\tau}^2<0$        & 0.92 & 0.13 & 0.043 & 23.1 & 75.6 & 12.8 & 7.7 
& 3.2 \\
$C1+angle$        & 0.68 & 0.12 & 0.04 & 1.8 & 130 & 22 & 11.0 & 4.9 \\
$C2$        & 0.29 & 0.049 & 0.019 & 0.088 & 0.99 & 0.49 & 0.18 & 0.14 \\
$C3$        & 0.25 & 0.032 & 0.017 & 0.088 & 0.29 & 0.39 & 0.15 & 0.07 \\
\hline
\end{tabular}
\caption{Cross sections (in $fb$) for signal benchmark points and the
  various SM backgrounds listed in the text after various cuts. The row
  labelled BC denotes parton level cross sections after  the requirement
  $p_T(j)> 80$~GeV, along with
minimal cuts
  implemented to regulate divergences, and also includes the leptonic
  branching fractions for decays of both the top quarks in the
  $t\bar{t}$ column. The remaining rows list the cross sections after a
  series of analysis cuts detailed in the text. }
\label{tab:xsec}
\end{table}
At this stage, the leading irreducible physics background from
$\tau\bar{\tau} j $ events is two orders of magnitude larger than the
signal, while the background from top pair production (which nominally
contains two untagged $b$-jets) is a factor of two larger than this.
\subsection{Reducing the di-tau plus jet background}

\subsubsection{$m_{\tau\tau}^2$ cut}
After {\bf C1} cuts, the decay $Z\to \tau\bar{\tau}$ of an on-shell high
$p_T$ $Z$ boson recoiling against a hard QCD jet is the leading source
of the $\tau\bar{\tau}j$ events. This means that this background can be
greatly reduced if it would be possible to reconstruct the di-tau
invariant mass {\em despite the presence of the neutrinos that are
  present when the taus decay leptonically.} This is possible because
the taus from $Z$-boson decays are typically very relativistic.  In the
approximation that the leptons and neutrinos from the decay of a
relativistic tau are all exactly collimated along the parent tau
direction, we can write the momentum carried off by the two neutrinos
from the decay $\tau_1\to \ell_1\bar{\nu}_{\ell_1}\nu_{\tau_1}$ of the
first tau as $\xi_1\vec{p}(\ell_1)$ and, similarly, as
$\xi_2\vec{p}(\ell_2)$ for the second tau.  Momentum conservation in the
plane transverse to the beams then requires that \be
-\sum_{jets}\vec{p}_T(j)=(1+\xi_1)\vec{p}_T(\ell_1
)+(1+\xi_2)\vec{p}_T(\ell_2 ) .
\label{eq:jetsum}
\ee 
These two equations can be solved for $\xi_1$ and $\xi_2$ given that
$\vec{p}_T(j)$ and $\vec{p}_T(\ell_{1,2})$ are all measured, and used to
evaluate the momenta of the individual taus.  This then allows us to
evaluate the invariant mass squared of the di-tau system which (within
the collinear approximation for tau decays) is given by,
\be
m_{\tau\tau}^2=(1+\xi_1)(1+\xi_2)m_{\ell\ell}^2 .
\label{eq:mtautau}
\ee
We show the distribution of $m_{\tau\tau}^2$ constructed using
Eq.~(\ref{eq:mtautau}) for both signal events as well as for the various
backgrounds in Fig. \ref{fig:mtt} after the cut set {\bf C1} and also
requiring $n_j=1$ to further reduce the top background.  As expected,
this peaks sharply around $m_Z^2$ for the $\tau\bar{\tau}j$ events (red
histogram). In contrast, for signal and other SM background sources,
where the isolated lepton and $\vec{\eslt}$ directions are uncorrelated,
the $m_{\tau\tau}^2$ distributions are very broad and peak at even
negative values of $m_{\tau\tau}^2$.  The $m_{\tau\tau}^2$ variable
provides a very good discriminator between $\tau\bar{\tau}j$ background
and signal. The cut $m_{\tau\tau}^2 < 0$ \cite{bmt} has, in fact,
been used in ATLAS \cite{atlas} and CMS \cite{cms} for their
analyses. We see from the third row of Table~\ref{tab:xsec} that the
$\tau\bar{\tau} j$ background is reduced by a factor $\sim 4$ while the signal
is reduced by 25-40\%. The efficiency with which the di-tau background
is reduced is limited because the tail of the $\tau\bar{\tau}j$
background extends out to negative values of $m_{\tau\tau}^2$; this happens
because of tau pair production from virtual photons, the breakdown of
the collinear approximation for asymmetric $Z$ decays and finally
hadronic energy mismeasurements which skew the direction of both
$\vec{p}_T(j)$ and of $\ \vec{\eslt}$.
\begin{figure}[!htbp]
\begin{center}
\includegraphics[height=0.4\textheight]{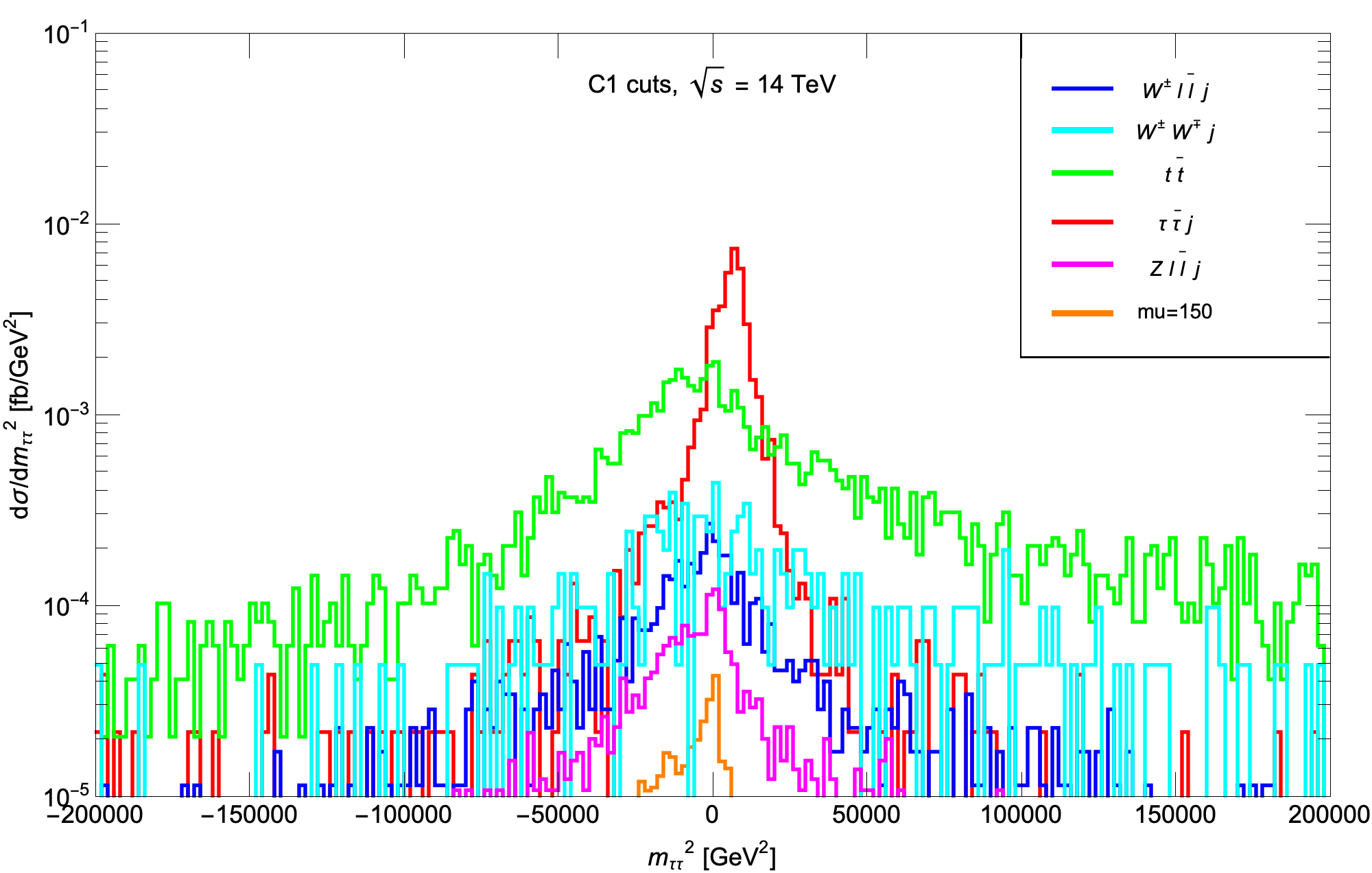}
\caption{Distribution in $m_{\tau\tau}^2$ for the three SUSY BM models
  with $\mu =150,\ 200$ and 300 GeV introduced in the text, along with
  SM backgrounds after $C1$ cuts augmented by $n_j=1$.
\label{fig:mtt}}
\end{center}
\end{figure}
In Sec.\ref{subsubsec:angle}, we describe angle cuts which can
reduce the di-tau background very efficiently with only moderate loss of
the higgsino signal, and suggest that these replace the $m_{\tau\tau}^2<
0$ cut that has been used by ATLAS and CMS for their compressed higgsino
search.

\subsubsection{ Angle cuts} \label{subsubsec:angle}

In the transverse plane, the di-tau pair must recoil against the hard
QCD radiation with the transverse plane opening angle between the taus
significantly smaller than $\pi$. The central idea behind our proposal
is that, in the transverse plane, the $\eslt$ vector {\it must lie
  between the directions of the two taus}. For relativistic taus, the
tau direction is, of course, essentially the same as the {\it
  observable} direction of its charged lepton daughter as
illustrated in Fig.~\ref{fig:sketch}. We require the azimuthal angles
$\phi_\ell$ and $\phi_{\bar{\ell}}$ for each lepton to lie between $0$
and $2\pi$, and define $\phi_{max}=max(\phi_\ell,\phi_{\bar{\ell}})$ and
$\phi_{min}=min(\phi_\ell,\phi_{\bar{\ell}})$. Then for $\vec{\eslt}$ to
lie in between the tau daughter lepton directions we must have,
\footnote{This works as long as $|\phi_{\ell}-\phi_{\bar{\ell}}|< \pi$. If
  $|\phi_{\ell}-\phi_{\bar{\ell}}|> \pi$, define $\phi_{\ell}^\prime
  =\phi_{\ell}+\pi$, $\phi_{\bar{\ell}}^\prime= \phi_{\bar{\ell}}+\pi$ and
  $\phi_{\hspace{1mm} \eslt}^\prime =\phi_{\hspace{1mm} \eslt}+\pi$, (all
  modulo $2\pi$) along with
  $\phi_{max}=max(\phi_\ell^\prime,\phi_{\bar{\ell}}^\prime)$, and likewise,
  $\phi_{min}=min(\phi_\ell^\prime,\phi_{\bar{\ell}}^\prime)$, and then require,
  $\phi_{min} < \phi_{\hspace{1mm} \eslt} < \phi_{max}$.}
$$\phi_{min} < \phi_{\hspace{1mm} \eslt}<\phi_{max}.$$ Notice that, by
definition, $\phi_{max}-\phi_{min} < \pi$, and for a boosted tau pair,
often significantly smaller than $\pi$.

\begin{figure}[!htbp]
\begin{center}
\includegraphics[height=0.23\textheight]{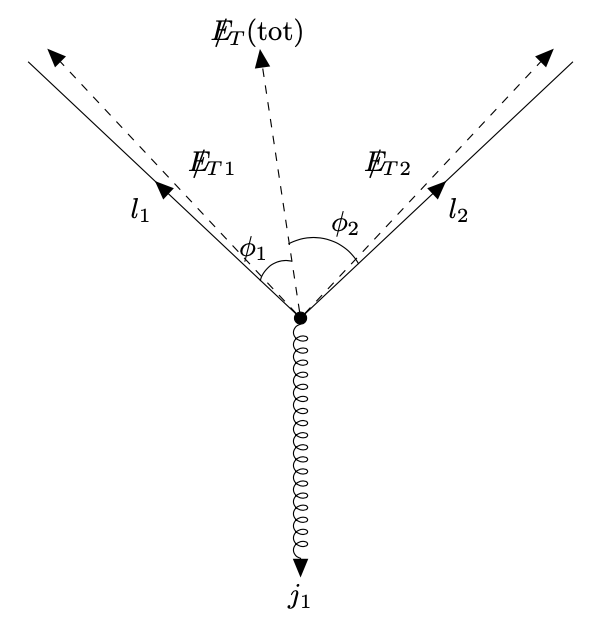}
\caption{Sketch of a ditau background event to the di-lepton plus jet plus
$\eslt$ signature in the transverse plane of the event. Here $\ell_1$ and
  $\eslt_{1}$ denote the transverse momentum of the lepton and of the
  vector sum of the neutrinos from the decay of the first tau, and
  likewise $\ell_2$ and $\eslt_2$. $\eslt$(tot) is the resultant $\eslt$
  in the event. Notice that because the taus are expected to be
  relativistic, $\ell_i$ and $\eslt_i$ vectors are nearly collimated
  along the direction of the $i^{th}$ tau ($i=1,2$). 
\label{fig:sketch}}
\end{center}
\end{figure}
To characterize the $Z(\to \tau \bar{\tau})+j$ background, we show in
Fig.~\ref{fig:phi1phi2} a scatter plot of these events in the
$\phi_1\equiv \phi_{max}-\phi_{\hspace{1mm} \eslt}$ vs. $\phi_2\equiv
\phi_{\hspace{1mm} \eslt}-\phi_{min}$ plane. If the collinear approximation 
for tau decays holds, we would expect that the $\tau\bar{\tau}
j$ background selectively populates the top right quadrant with
$\phi_1>0$ and $\phi_2>0$ with $\phi_1+\phi_2 =\phi_{max}-\phi_{min} <
\pi$, and significantly smaller than $\pi$ when the tau pair emerges
with a small opening angle in the transverse plane. We see from the
figure that there is a small, but significant, ``spill-over'' into the
region where $\phi_1$ or $\phi_2$ assumes small negative values; {\it
  i.e.} where $\vec{\eslt}$ lies just outside the cone formed by
$\vec{\ell_1}$ and $\vec{\ell_2}$. This spill-over arises from
asymmetric decays of the $Z$ where one of the taus (the one emitted
backwards from the $Z$ direction) is not as relativistic so
that the collinear approximation works rather poorly, or because hadronic
energy mismeasurements skew the direction of $\vec{\eslt}$.
Indeed we see from the top frame of Fig.~\ref{fig:phi1phi2} that the
$\tau\bar{\tau} j$ background mostly populates the triangle in the top-right
quadrant of the $\phi_1$ vs. $\phi_2$ plane, and $\phi_1+\phi_2< f\pi$
where the fraction $0<f<1$, with a spill-over into the strips where one
of $\phi_{1,2}$ is slightly negative.  For signal events and for the
other backgrounds, $\phi_{\hspace{1mm} \eslt}$ will be uncorrelated with
$\phi_{min}$ and $\phi_{max}$, and so their scatter plots will extend to
the other quadrants. This is illustrated for the $t\bar{t}$ background
in the middle frame of Fig.~\ref{fig:phi1phi2} and for signal point BM1 in
bottom frame. In these cases, we indeed see a wide
spread in $\phi_1$ and $\phi_2$ between $\pm 2\pi$.

\begin{figure}[!htbp]
\begin{center}
\includegraphics[height=0.3\textheight]{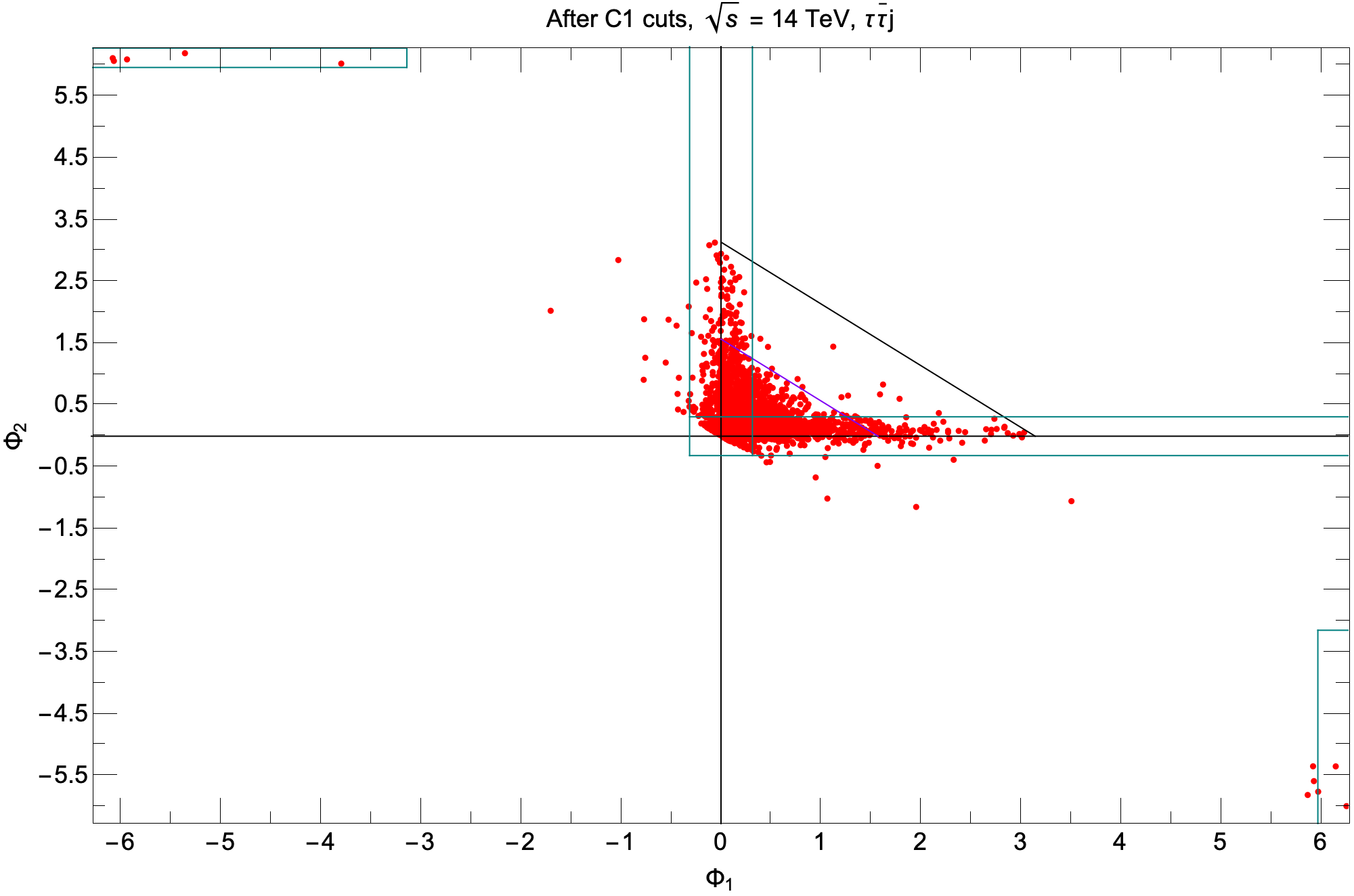}\\
\includegraphics[height=0.3\textheight]{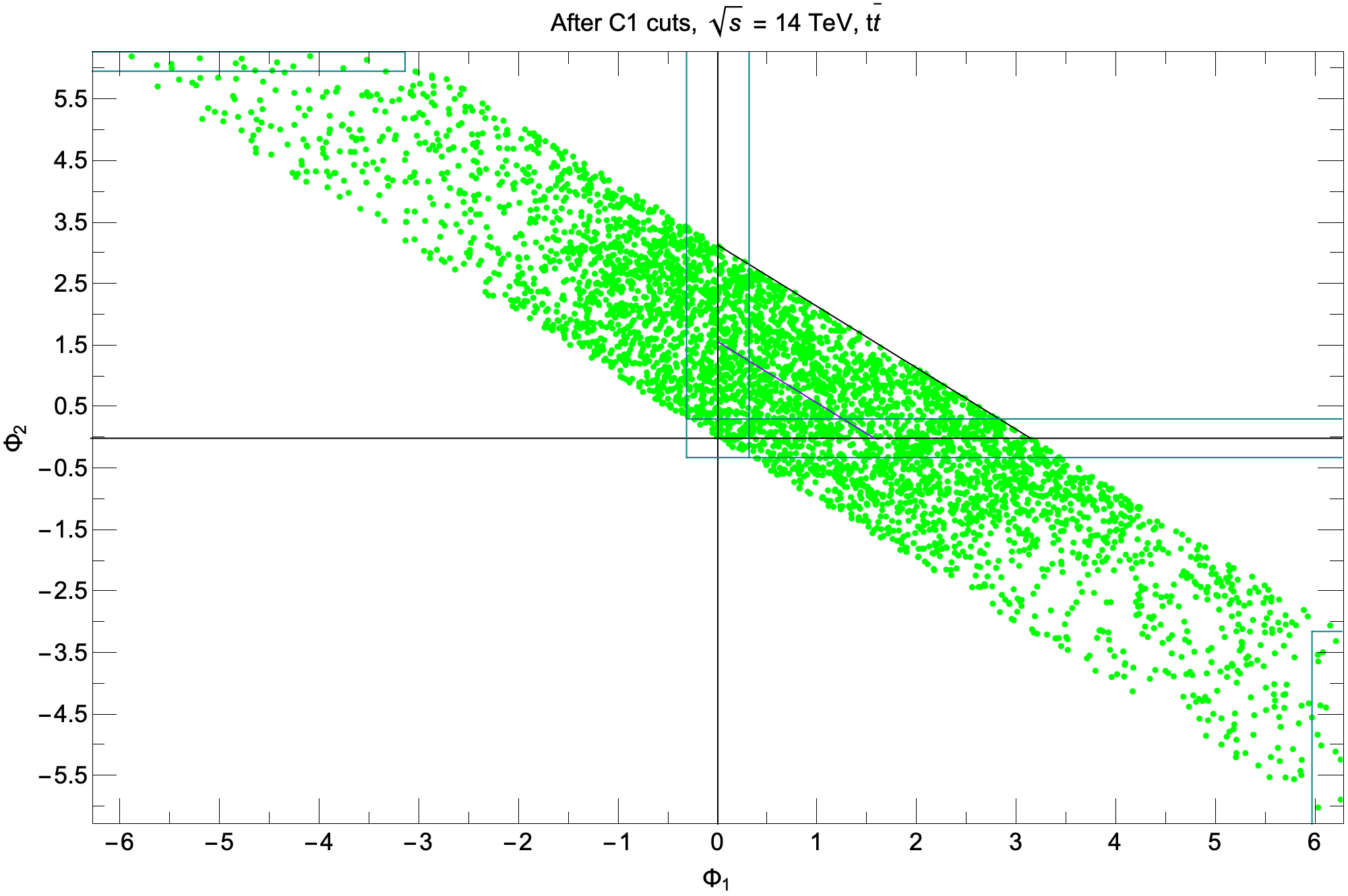}\\
\includegraphics[height=0.3\textheight]{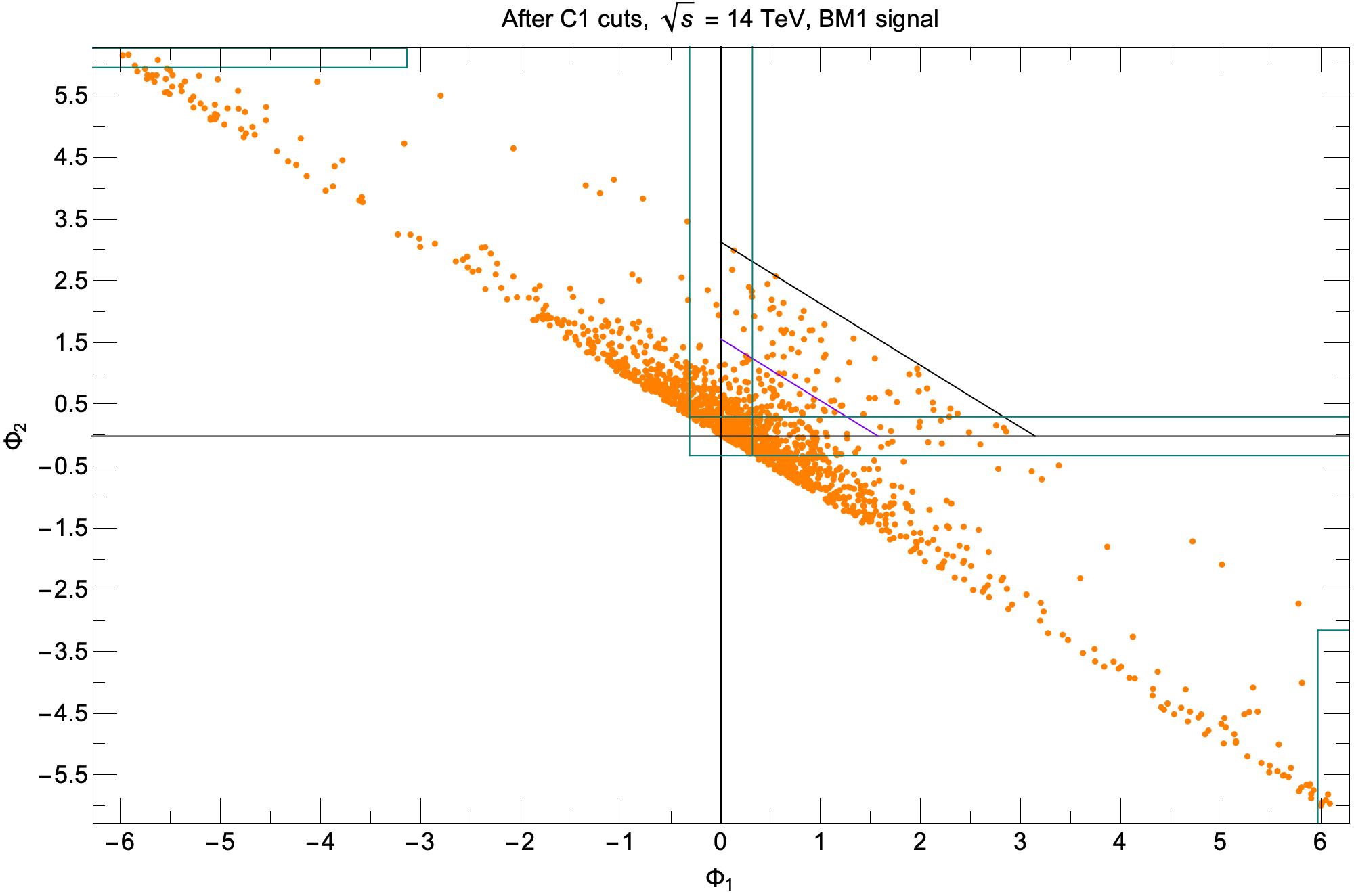}
\caption{Scatter plots in the $\phi_1$ vs. $\phi_2$ plane for
  $\tau\bar{\tau}j$ and $t\bar{t}$ backgrounds and for the signal point
  BM1 after $C1$ cuts, requiring also that $n_j=1$.  Also shown are the
  lines where $\phi_1+\phi_2 < \pi/2, \pi$ along with the strip cuts in
  Eq.~(\ref{eq:strip}).  In the strips in the top left and bottom right
  of each frame and along the positive $\phi_{1}$ and $\phi_2$ axes,
  $\ \vec{\eslt}$ is close to the boundary of the lepton cone.
\label{fig:phi1phi2}}
\end{center}
\end{figure}

To efficiently veto the $\tau\bar{\tau}j$ background, we examined nine
cases of angular cuts \cite{angular}. Specifically, we allowed $f=1,2/3$
and 1/2 to optimize the effect of the boost of the di-tau on the
transverse plane opening angle, and also allowed for the width of
the spill-over strip (to negative values of $\phi_1$ or $\phi_2$) equal
to 0, $\pm \pi/20$ and $\pm \pi/10$.  We found that $S/\sqrt{BG(\tau\bar{\tau}
  j)}$ was optimized for LHC14 with 3000 fb$^{-1}$ when we,
\be {\rm veto\ the\ triangle}\ \ \ \phi_1,\ \phi_2> 0\ \ {\rm
  with}\ \ \phi_1+\phi_2<\pi /2, \label{eq:phi}
\ee 
along with an additional veto 
to further
reduce background from the spill-over of $\ \vec{\eslt}$ outside of the
cone defined by the charged leptons:
\be
{\rm strip\ cuts:\ veto} \ |\phi_{1,2}|<\pi /10 , \ {\rm and}\ 
{\rm veto}
\ \phi_{1,2} \ {\rm in \ the \ two \ corner \ strips.}
\label{eq:strip}
\ee
In the horizontal (vertical) strip in the upper left (lower right)
quadrant of each frame of Fig.~\ref{fig:phi1phi2}, $\phi_2$ ($\phi_1$)
is within $\pi/10$ of $2\pi$ which, of course, is the same $-\pi/10 <
\phi_{1,2} < 0$ (modulo $2\pi$), and hence just outside the lepton cone.

We list signal and background rates after {\bf C1} cuts together with
the cuts (\ref{eq:phi}) and (\ref{eq:strip}) -- collectively referred to
as {\em angle cuts} from this point on -- in row 4 of
Table~\ref{tab:xsec}. In this case, we find that $\tau\bar{\tau}j$
background is reduced from cut set {\bf C1} by a factor $\sim 52$
(compared to a factor $\sim 4$ for the $m_{\tau\tau}^2< 0$ cut) whilst
signal efficiency for the point BM1 is almost 60\% (compared to $\sim
75$\% for the $m_{\tau\tau}^2< 0$ cut).  We also see that signal
efficiency for the other two benchmark points is nearly the same for the
angular and for the $m_{\tau\tau}^2< 0$ cuts.  We regard the angular
cuts as a significantly improved method for almost completely removing
the irreducible $\tau\bar{\tau}j$ background. The other SM backgrounds
are not as efficiently reduced by the angular cut as by the
$m_{\tau\tau}^2< 0$ cut, and need other cuts to reduce them to
manageable levels.

\subsection{Further cuts to remove top and $W$-pair backgrounds} \label{subsec:addcuts}

We have just seen that while the angle cuts greatly reduce the
irreducible $\tau\bar{\tau} j$ background, SM backgrounds from
$t\bar{t}$ and $WWj$ events (followed by leptonic decays of the top and
$W$-bosons) and also from $W/Z\ell\bar{\ell}+j$ production still
completely dwarf the signal. Since the top pair background typically has
a higher jet multiplicity, requiring $n_j=1$ increases the
signal-to-background ratio (as already mentioned earlier). The hard jet
and $\eslt$ distributions are both backed up against their cut value,
and given the already small signal cross section, it is not helpful to
require a harder cut on these variables. 

In contrast, the transverse momenta of the leptons from 
on-shell top and $W$ decays then have broader distributions than the
signal leptons. We found that requiring upper limits on
$|p_T(\ell_2)|$ and $H_T(\ell\bar{\ell})\equiv
|p_T(\ell_1)|+|p_T(\ell_2)|$ indeed enhances the signal over the
background \cite{angular}. The transverse momenta of leptons and
neutrinos from $W$ and top decays have comparable magnitudes; in
contrast the momentum scale for signal leptons is set by $\Delta m$
while that for signal $\ \eslt$ is set by the higgsino mass. As a result,
the ratio $\ \eslt /H_T(\ell\bar{\ell})$ is expected to be considerably
harder for the higgsino signal than for the SM background as first noted
by the ATLAS collaboration \cite{atlas1}. Finally, we note that the
dilepton mass distribution for leptons from $\tz_2\to
\tz_1\ell\bar{\ell}$ decays is kinematically bounded by
$m(\ell\bar{\ell})< m_{\tz_2}-m_{\tz_1} \alt 20-25$~GeV for the
compressed higgsino search, and further, that the invariant mass of
dileptons from $\tchi_1^+\tchi_1^-$ production events will also tend to be
smaller than for background events, simply because leptons from the
chargino decays are soft. Again, we refer the interested reader to
Ref.\cite{angular} where these distributions and others are explicitly
shown.

These considerations lead us to include the analysis cut set {\bf C2} to
enhance the higgsino signal over the top and $WWj$
backgrounds that dominate after the angle cuts:
\bi
\item the cut set {\bf C1} together with the ${\rm angle\ cuts}$,
\item $n(jets) = 1$,
\item $p_T(\ell_2 ):5-15$ GeV,
\item $H_T(\ell\bar{\ell})<60$ GeV
\item $\eslt/H_T(\ell\bar{\ell} )>4$, and
\item $m(\ell\bar{\ell})<50$ GeV.
\ei 
We have checked that the requirement $m(\ell\bar{\ell})<
50$~GeV, efficiently reduces much of the background while retaining most
of the higgsino signal.

We see from the penultimate row of Table~\ref{tab:xsec} that after {\bf
  C2} cuts, the leading $t\bar{t}$ background has dropped by a factor
$\sim 130$, and the total SM background has dropped to $\sim 1.1$\%,
while the signal is retained with an efficiency of 40-60\%.  At this
point, the total background is just below 2~fb.  Clearly, the signal
cross section is small, and the large integrated luminosities expected
at the HL-LHC will be necessary for the detection of the signal if the
higgsino mass is close to its naturalness bound of 300-350~GeV, or if
the higgsino spectrum is maximally compressed to the $4-5$ GeV level,
consistent with electroweak naturalness.

To further enhance the signal relative to (particularly the top)
background, we note that for the signal, where the SUSY particles recoil
strongly against the ISR jet, we expect nearly back-to-back
$\vec{p}_T(jet)$ and $\vec{\eslt}$ vectors.  This correlation is
expected to be somewhat weaker from  $t\bar{t}$ background events and also
for $W\ell\bar{\ell}j$ events
because these intrinsically contain additional activity
from decay products that do not form jets or identified leptons.  
The dilepton-plus-$\eslt$ cluster transverse
mass $m_{cT}(\ell\bar{\ell},\eslt)$ and the relative values of
$E_T(jet)$ and $\eslt$ (which tend to be more balanced for the signal
than for the backgrounds) serve to give added distinction between the
signal and backgrounds and provide additional discriminators. 
These considerations led us to impose the additional
cut set {\bf C3} that includes:
\bi
\item all {\bf C2} cuts, 
\item $\Delta\phi (j_1,\eslt )>2.0$
\item $m_{cT}(\ell\bar{\ell} ,\eslt)<100$ GeV
\item $p_T(j_1)/\eslt <1.5$
\item $|p_T(j_1)-\eslt |<100$ GeV. 
\ei 
                     
We show the OS/SF dilepton invariant mass after these {\bf C3} cuts
in Fig. \ref{fig:mllC3}.  The total background, shown in gray, is
essentially flat, whereas signal-plus-background is shown by the colored
histogram, and correspond to {\it a}) BM1 with $\Delta m=12$ GeV, {\it
  b}) BM2 with $\Delta m=16$ GeV and {\it c}) BM3 with $\Delta m=4.3$
GeV.  The idea here is to look for systematic deviations from SM
background predictions in the lowest $m(\ell\bar{\ell})$ bins.  Those
bins with a notable excess could determine the kinematic limit
$m(\ell\bar{\ell})<m_{\tz_2}-m_{\tz_1}$. By taking only the bins with a
notable excess, {\it i.e.}  $m(\ell\bar{\ell})<m_{\tz_2}-m_{\tz_1}$,
then it is possible to compute the cut-and-count excess above expected
background to determine a discovery limit or exclusion bound. The {\em
  shape} of the distribution of the excess below the $\tz_2 \to
\tz_1\ell\bar{\ell}$ end point depends on the {\em relative sign} of the
lighter neutralino eigenvalues (these have opposite signs for higginos)
and so could serve to check the consistency of higgsinos as the origin
of the signal\cite{shape}. Of the three cases shown, this would be
possible at the HL-LHC only for the point BM1, since the tiny signal to
background ratio precludes the possibility of determining the signal
shape in the other two cases. It may be worthwhile to examine whether
sophisticated machine learning methods could lead to a better signal and
background discrimination that allows us to obtain the neutral higgsino
mass gap and also the signal shape, particularly for the two difficult cases in
Fig.~\ref{fig:mllC3}
\begin{figure}[!htbp]
\begin{center}
\includegraphics[height=0.28\textheight]{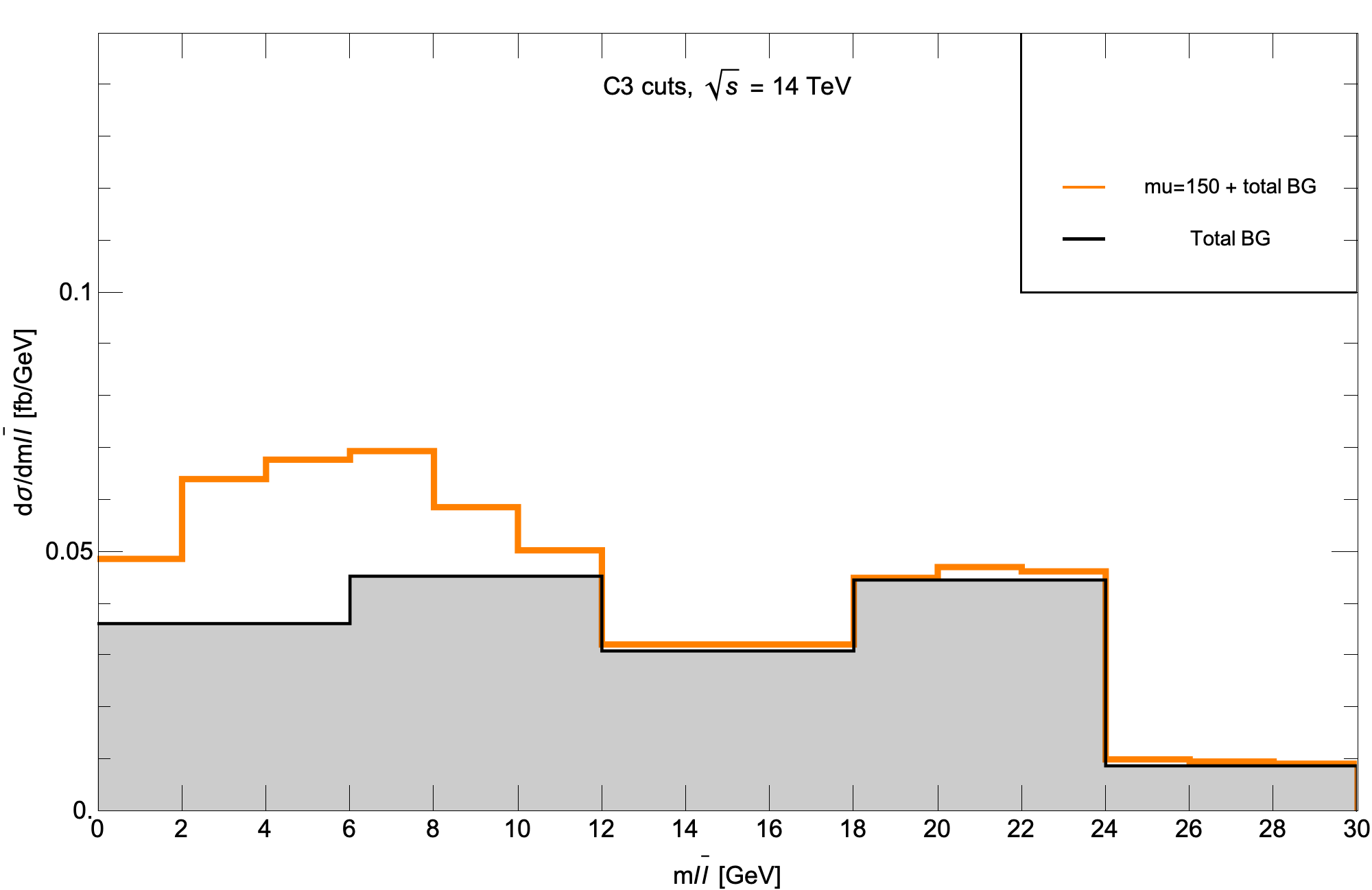}\\
\includegraphics[height=0.28\textheight]{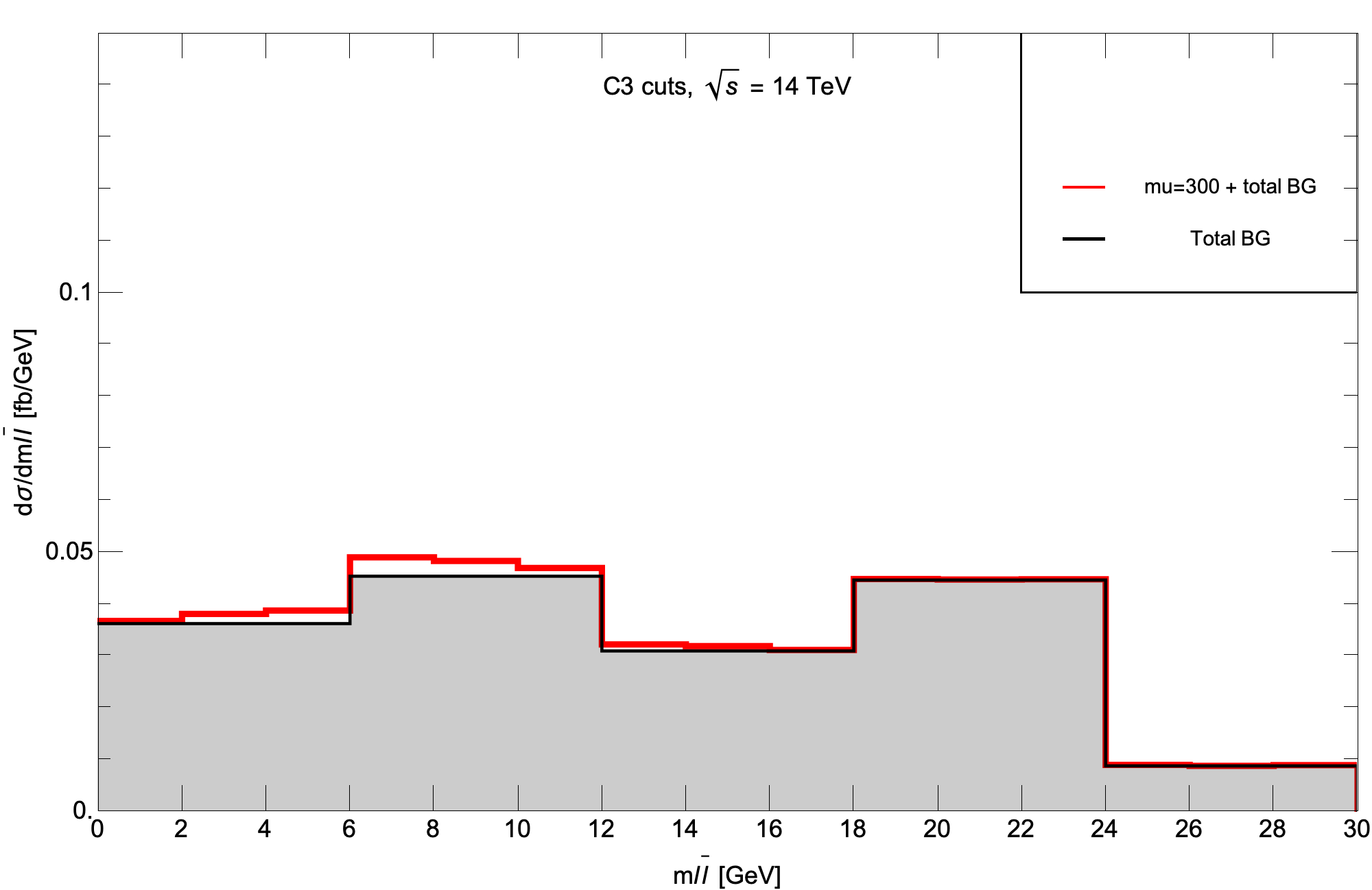}\\
\includegraphics[height=0.28\textheight]{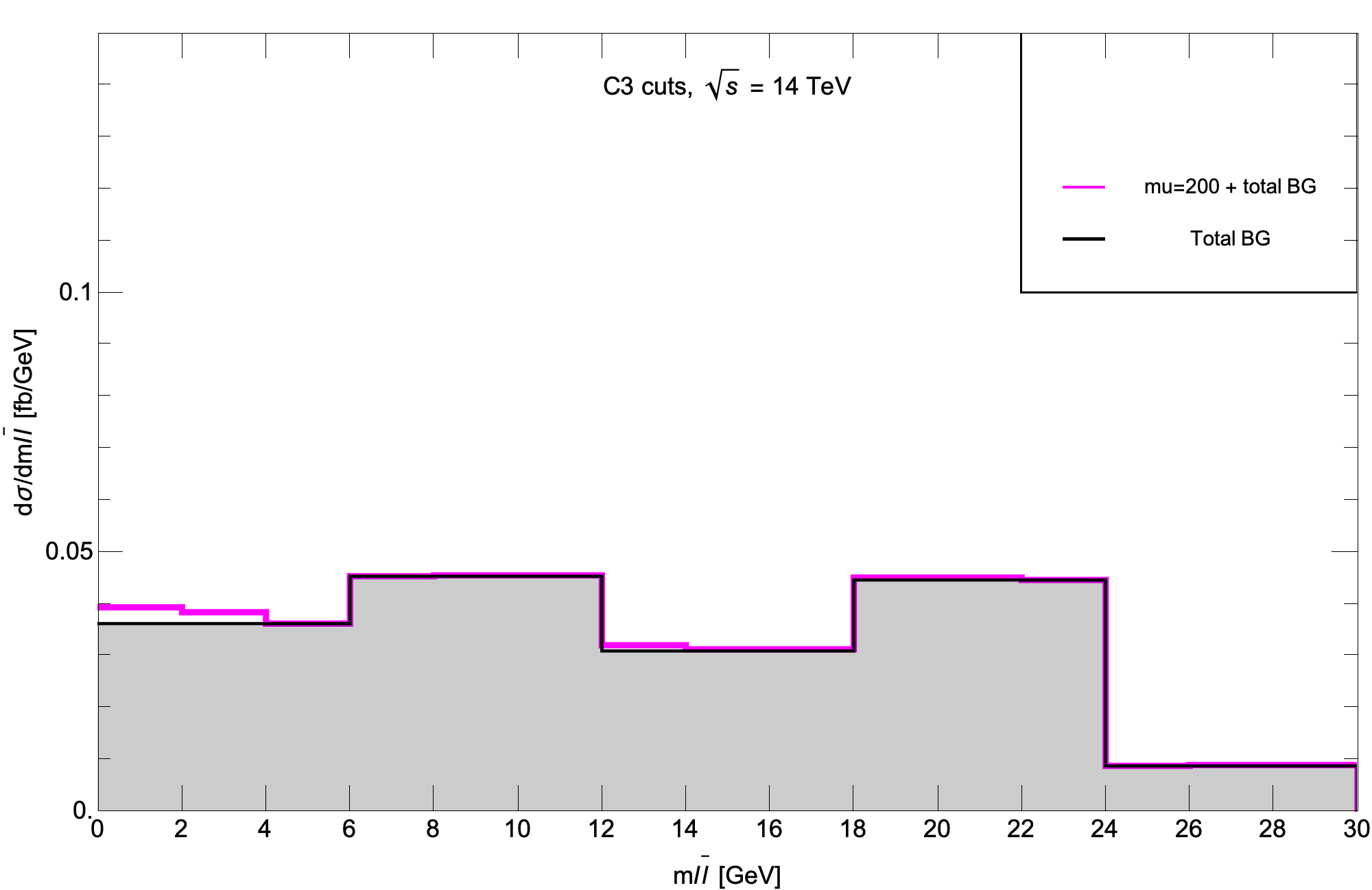}
\caption{Distribution of $m(\ell^+\ell^-)$ for the three SUSY BM models with 
$\mu =150,\ 300$ and 200 GeV, and for the  SM backgrounds after $C3$ cuts.
\label{fig:mllC3}}
\end{center}
\end{figure}

\section{LHC reach for higgsinos with 300-3000~fb$^{\bf {-1}}$}
\label{sec:reach}

The distributions in Fig.~\ref{fig:mllC3} suggest that 
our final analysis cut set {\bf C4} include,
\bi
\item {\bf C3} cuts, along with
\item $m(\ell\bar{\ell})<m_{\tz_2}-m_{\tz_1}$.
\ei

The location of the mass gap is clearly visible for BM1, but is
obscured by the background for the other two cases. To implement the
last $m(\ell\bar{\ell})$ cut, we suggest examining the cross section
with $m_{\ell\ell}< m_{\ell\ell}^{\rm cut}$ for varying the value of
$m_{\ell\ell}^{\rm cut}$ and looking for a rise in the event rate where
events from $\tz_2\to\tz_1\ell\bar{\ell}$ would be expected to
accumulate. We recognize that with the cut-and-count technique this may
be not be possible for the difficult cases BM2 and BM3, but in the
following, we will optimistically assume that once we have the data, the
region where the higgsino signal is beginning to accumulate will be
self-evident.
\begin{figure}[!htbp]
\centering
\begin{subfigure}[t]{0.4\textwidth}
  \centering
  \includegraphics[width=1.1\linewidth]{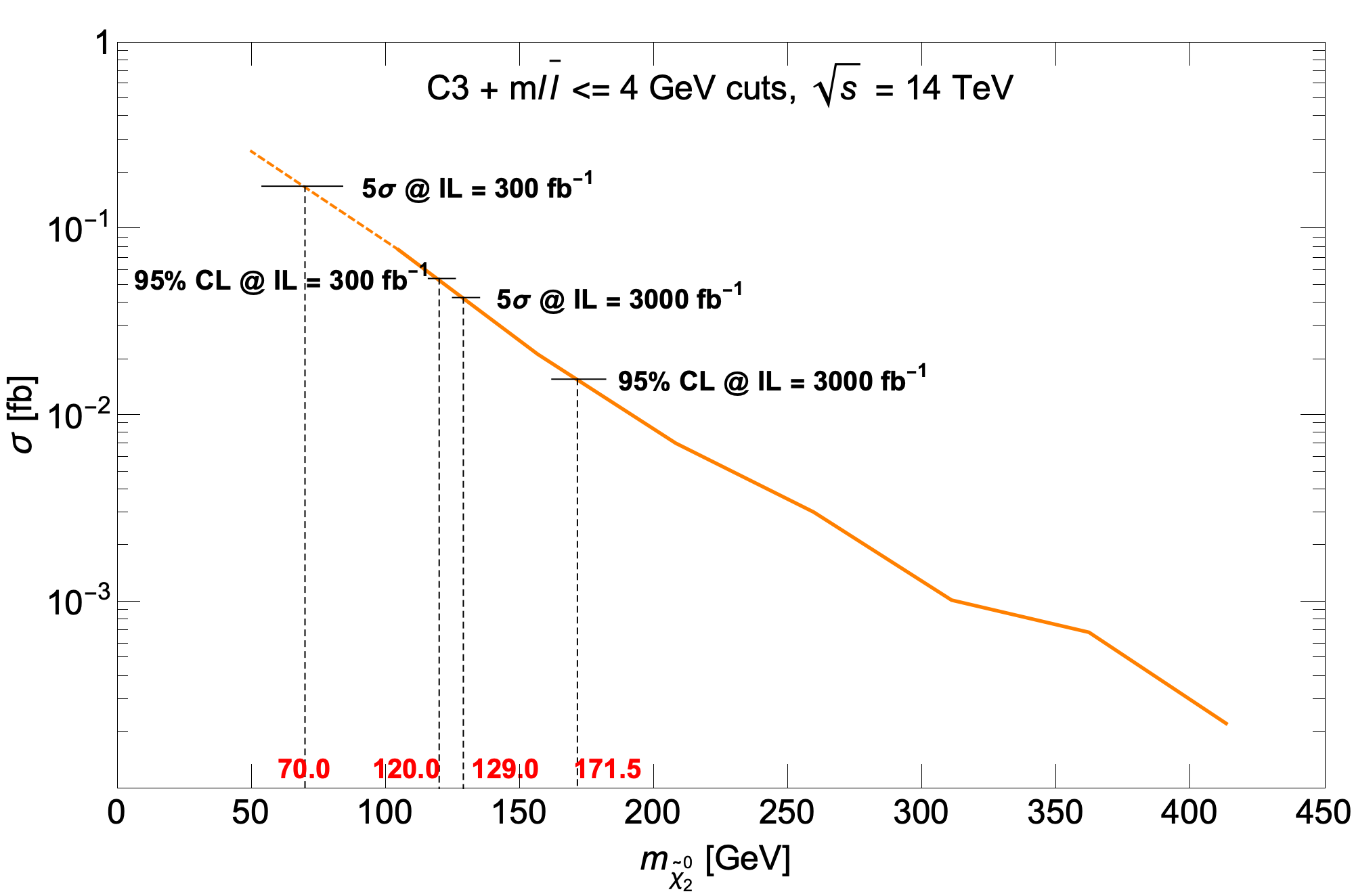}
  \caption{}
  \label{fig:mdlln_4}
\end{subfigure}%
\quad \quad
\begin{subfigure}[t]{0.4\textwidth}
  \centering
  \includegraphics[width=1.1\linewidth]{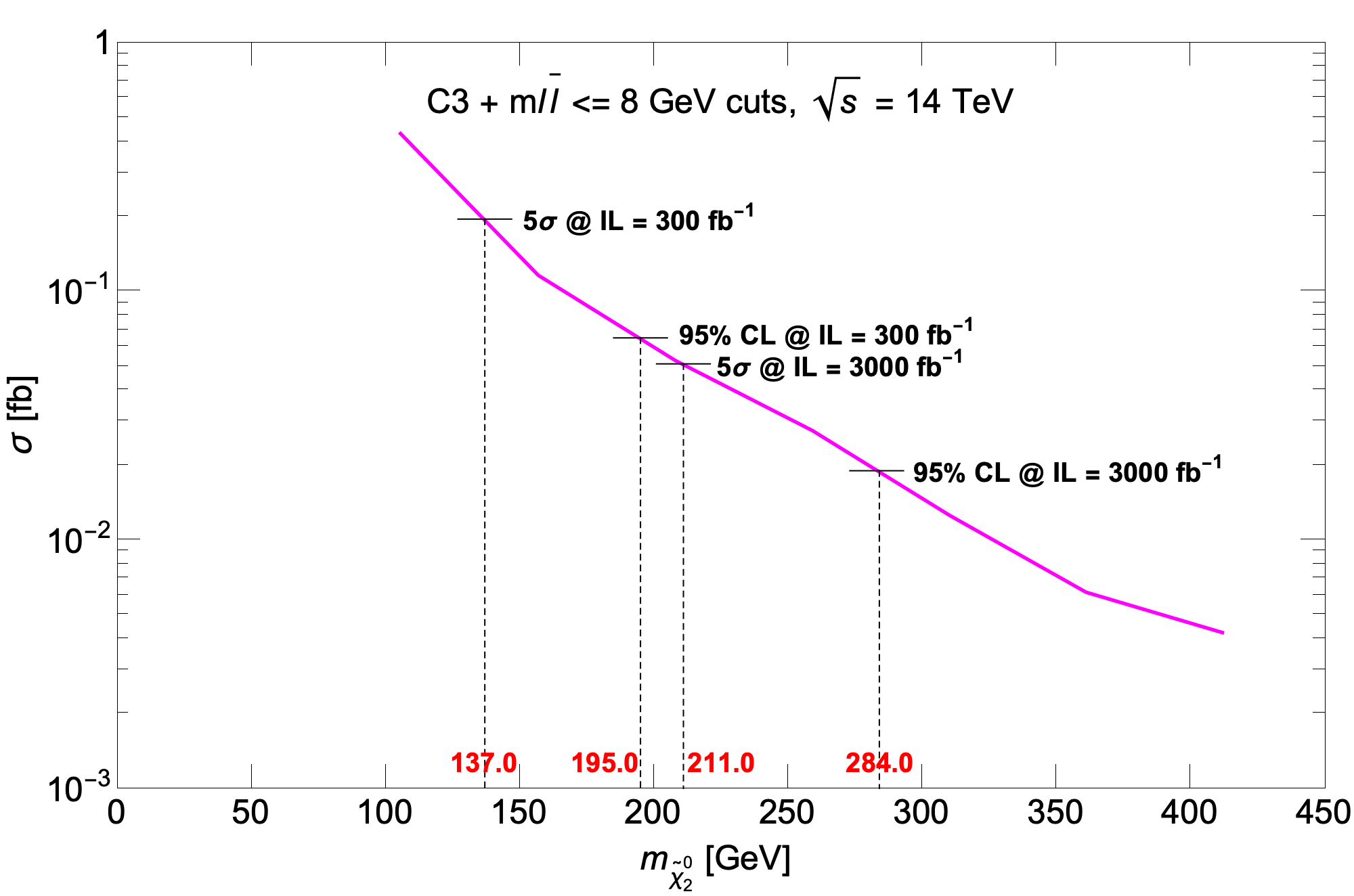}
  \caption{}
  \label{fig:mdlln_8}
\end{subfigure}
\begin{subfigure}[t]{0.4\textwidth}
  \centering
  \includegraphics[width=1.1\linewidth]{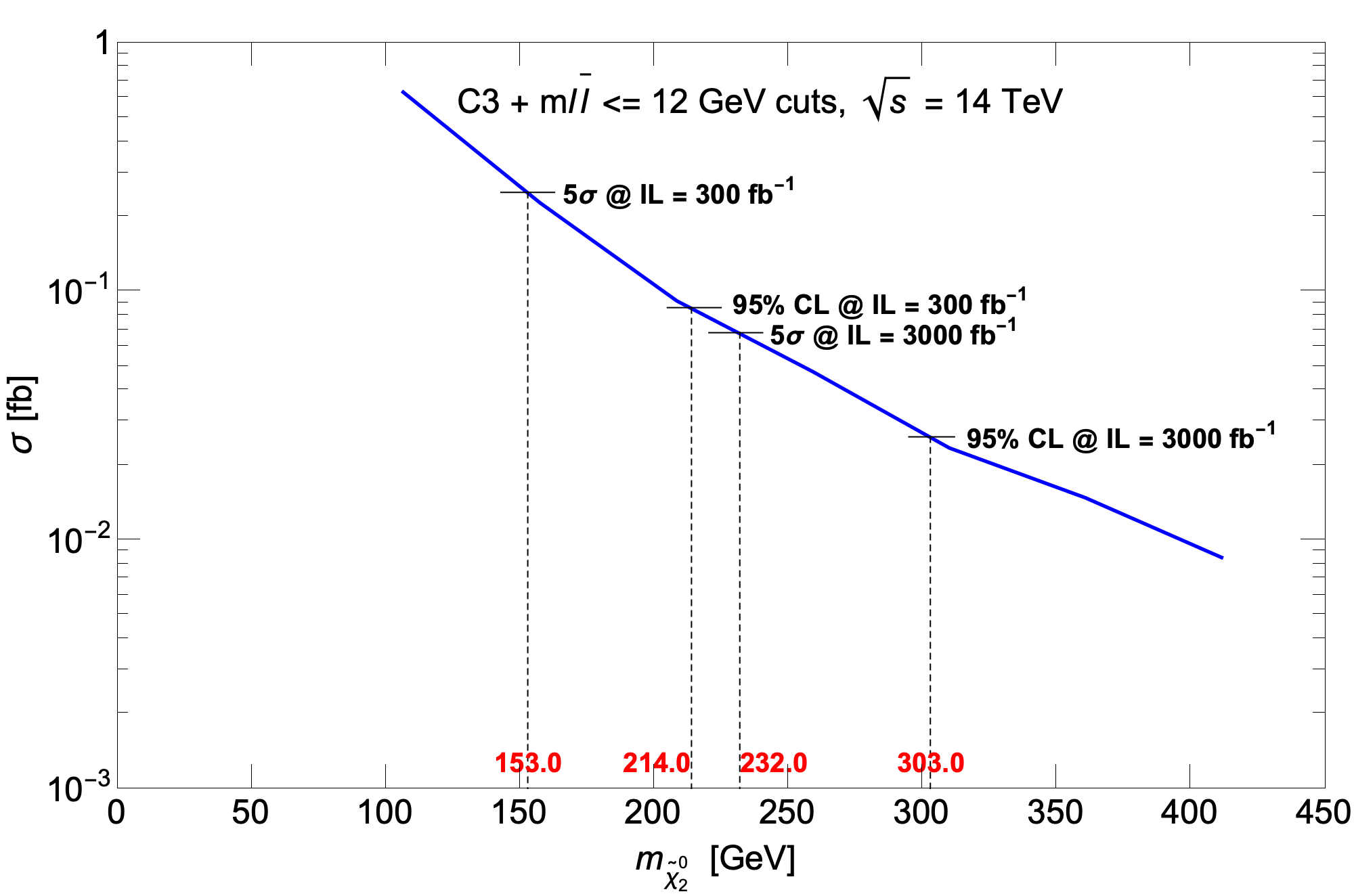}
  \caption{}
  \label{fig:mdlln_12}
\end{subfigure}%
\quad \quad
\begin{subfigure}[t]{0.4\textwidth}
  \centering
  \includegraphics[width=1.1\linewidth]{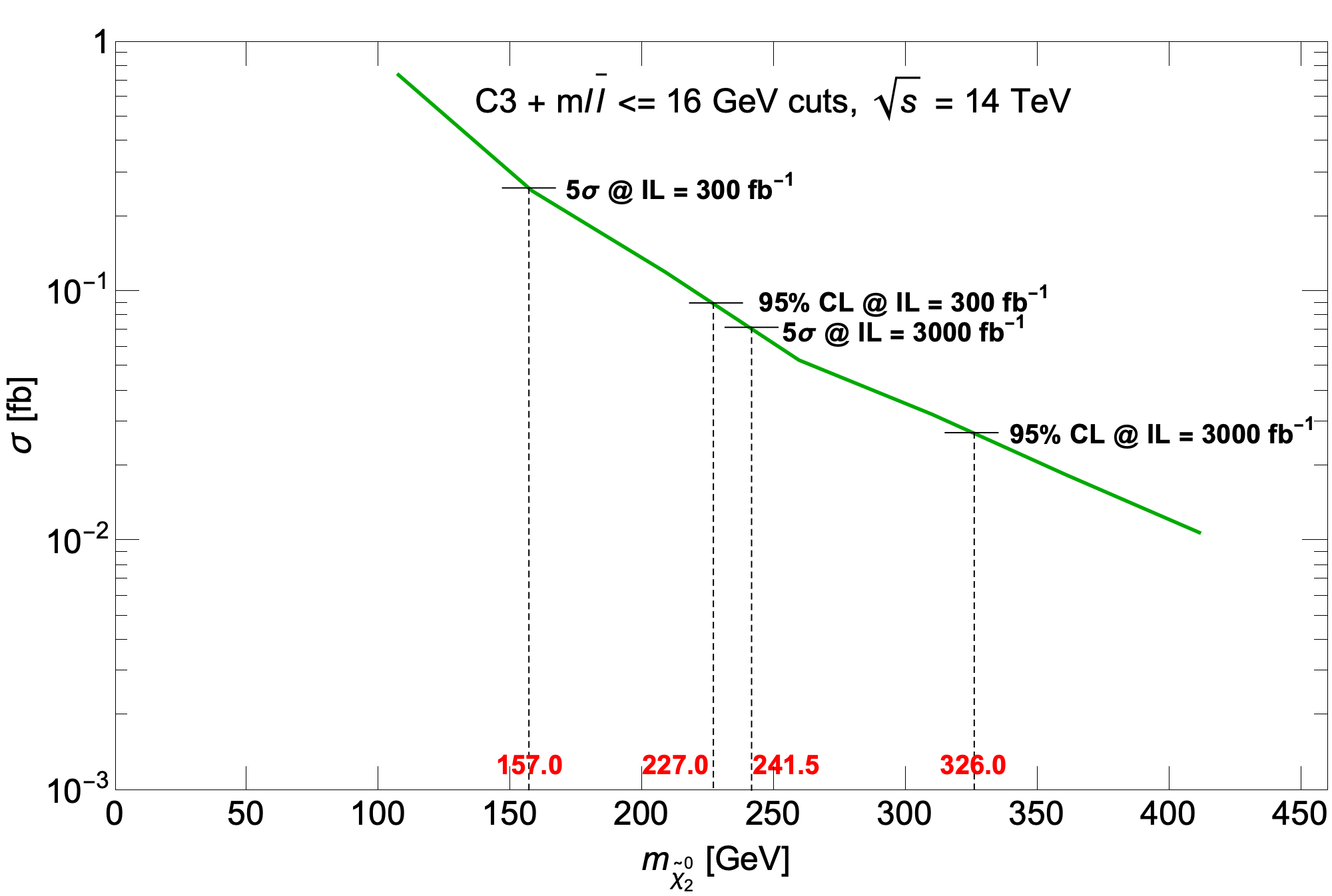}
  \caption{}
  \label{fig:mdlln_16}
\end{subfigure}
\caption{The projected $5\sigma$ reach and 95\% CL exclusion of the HL-LHC
  with 3000 fb$^{-1}$ in $\mu$ for four different NUHM2 model lines with
  {\it a}) $\Delta m=4$ GeV, {\it b}) $\Delta m=8$ GeV, {\it c}) $\Delta
  m=12$ GeV and {\it d}) $\Delta m=16$ GeV after $C3$ + $m(\ell\bar{\ell})<m_{\tz_2}-m_{\tz_1}$ cuts.
\label{fig:reachC4}}
\end{figure}

Using these {\bf C4} cuts, then we computed the signal cross section for
four model lines in the NUHM2 model for values of $\mu :100-400$ GeV and
with $m_{1/2}$ values adjusted such that the $m_{\tz_2}-m_{\tz_1}$ mass
gap is fixed at 4, 8, 12 and 16~GeV. While $\mu$ and $m_{1/2}$ are
variable, the values of $m_0=5$ TeV, $A_0=-1.6 m_0$, $\tan\beta =10$ and
$m_A=2$ TeV are fixed for all four model lines.  In
Fig. \ref{fig:reachC4}, we show the signal cross section after {\bf C4}
cuts, along with the $5\sigma$ reach and the 95\% CL exclusion for LHC14
with 300 and 3000 fb$^{-1}$. We see from Fig. \ref{fig:reachC4} that the
HL-LHC typically increases the higgsino reach by about 70-100~GeV
(50-60~GeV for the lowest values of $\Delta m=4$~GeV) as compared to the
reach that can be obtained with an integrated luminosity of
300~fb$^{-1}$.

\begin{figure}[!htbp]
\begin{center}
\includegraphics[height=0.4\textheight]{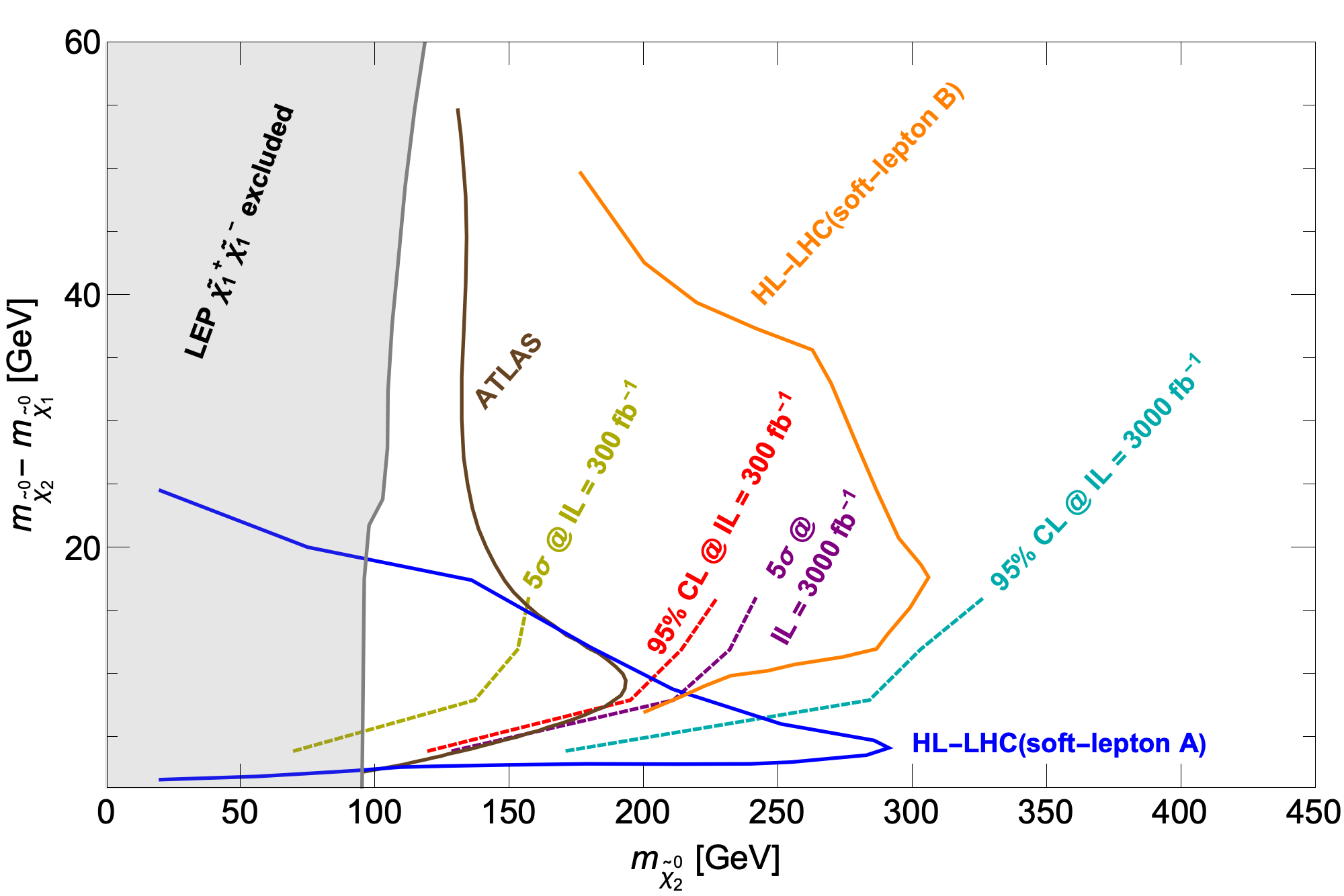}
\caption{The projected $5\sigma$ reach and 95\% CL exclusion contours
  for LHC14 with 300 and 3000 fb$^{-1}$ in the $m_{\tz_2}$ vs. $\Delta
  m$ plane after $C4$ cuts. Also shown is the current 95\% CL exclusion
  (ATLAS) and the projected 95\% CL exclusions from two different
  analyses for the HL-LHC \cite{Canepa:2020ntc}.
\label{fig:reach}}
\end{center}
\end{figure}
In Fig. \ref{fig:reach}, we translate the results of
Fig.~\ref{fig:reachC4} into contours in the $m_{\tz_2}$ vs. $\Delta m$
plane. The shaded region is excluded by LEP2 chargino searches.  The
region left of the contour labelled ATLAS is currently excluded by LHC
searches at the 95\%CL. We also show projections for what future
searches at the HL-LHC would probe at the 95\%CL \cite{Canepa:2020ntc}:
ATLAS (soft-lepton A) and CMS (soft-lepton B).  The various dashed
contours show our projections for the signal reach.  Our focus here has
been on higgsino mass gaps $\alt 20-25$~GeV, that are generically
expected in natural SUSY models. For larger mass gaps, it may be best to
search for higgsinos via the hard multilepton events, without any
requirement of a QCD jet. The reach projections in Fig.~\ref{fig:reach}
may be compared to theoretical expectations, both from bottom-up naturalness
considerations and from the string landscape \cite{Baer:2020sgm}.

Before closing this section, we note that we have only considered
physics backgrounds in our analysis. The ATLAS collaboration
\cite{atlas} has, however, reported that a significant portion of the
background comes from fake leptons, both $e$ and $\mu$. Accounting for
these detector-dependent backgrounds require data driven methods which
are beyond the scope of our analysis. We remark, however, that for any
specified value of the fake rate, it should be possible to make a rough
estimate of the impact of the fakes on the reach contours in
Fig.~\ref{fig:reach} using the cross sections in Fig.~\ref{fig:reachC4}
since if the fakes increase the background by a factor $f$, the cross
section necessary to maintain the same significance for the signal would
have to increase by $\sqrt{f}$.

\section{Summary}
\label{sec:summary}

Light higgsinos of mass $\sim 100-400$ GeV with a compressed spectrum
are the most robust prediction of
natural supersymmetry, and could be the only directly accessible
superpartners at the LHC. Here, we have re-examined the prospects for a
search for soft opposite-sign/same flavor dilepton plus $\eslt$ from
higgsino pair production in association with a hard monojet at the LHC.
The dileptons originating from $\tz_2\to\ell\bar{\ell}\tz_1$ would
exhibit a distinctive kinematic edge with
$m(\ell\bar{\ell})<m_{\tz_2}-m_{\tz_1}$, while the monojet and the
accompanying $\eslt$ serve as event triggers.

Our emphasis has been on the reduction of the main irreducible
background to the higgsino signal from $\tau\bar{\tau} j$ production
in the SM, where the soft leptons come from the decays of the taus. To
this end, we have proposed angular cuts (see Sec.~\ref{subsubsec:angle})
which appear to be more efficient in reducing this background than the
$m_{\tau\tau}^2$ cut that has been used by the ATLAS and CMS
collaborations in their analysis of the higgsino signal. Of course, additional
cuts are needed to further reduce the reducible background from
$t\bar{t}$ production, and subdominant backgrounds from $WWj$,
$W\ell\bar{\ell}j$ and $Z\ell\bar{\ell}j$ production.
Table~\ref{tab:xsec} shows a comparison between the new angular cuts and
the currently used di-tau mass cut, and also provides a cut flow after
various other analysis cuts discussed in the text.

Our final result is summarized by the dashed contours in
Fig.~\ref{fig:reach}. Our analysis works best for $\Delta m$ values
$\sim 15-20$ GeV but drops off for smaller and much larger mass gaps. We
mention that mass gaps smaller than about 4 GeV occur only for very
heavy gauginos that fail to satisfy naturalness expectations, while
higgsinos with an increasingly uncompressed spectrum can be more
effectively searched for via multilepton channels.  We see from
Fig.~\ref{fig:reach} that the HL-LHC with 3000 fb$^{-1}$ gives a
$5\sigma$ discovery reach to $m_{\tz_2}\sim 240$ GeV, with the 95\% CL
exclusion limit extending to $\sim 325$ GeV for $\Delta m\sim 16$
GeV. The signal reach would be even further enhanced if it becomes
possible to extend the lepton acceptance to lower $p_T$ values, or
reliably increase $b$-quark rejection even beyond 80-85\% that has
already been achieved.

{\it Acknowledgements:} 

This research is based upon work supported in part by the
U.S. Department of Energy, Office of Science, Office of Basic Energy
Sciences Energy Frontier Research Centers program under Award Number
DE-SC-0009956 and U.S. Department of Energy Grant DE-SC-0017647. The
work of DS was supported by the Ministry of Science and Technology
(MOST) of Taiwan under Grant No. 110-2811-M-002-574.


%
\end{document}